\documentclass[aps,superscriptaddress,showpacs,twocolumn]{revtex4}
\usepackage{hyperref}
\usepackage{graphicx}
\usepackage{bbm,amssymb}

\begin{document}
\title{On the topological conjugacy problem for interval maps}

\author{Roberto Venegeroles}\email{roberto.venegeroles@ufabc.edu.br}
\address{Centro de Matem\'atica, Computa\c c\~ao e Cogni\c c\~ao, UFABC, 09210-170, Santo Andr\'e, SP, Brazil}

\date{\today}

\begin{abstract}
{We propose an inverse approach for dealing with interval maps based on the manner whereby their branches are related (folding property), instead of addressing the map equations as a whole. As a main result, we provide a symmetry-breaking framework for determining topological conjugacy of interval maps, a well-known open problem in ergodic theory. Implications thereof for the spectrum and eigenfunctions of the Perron-Frobenius operator are also discussed.}
\end{abstract}


\maketitle

{\it Introduction} - From the ergodic point of view, a measure describing a chaotic map $T$ is rightly one that, after iterate a randomly chosen initial point, the iterates will be distributed according to this measure almost surely. This means that such a measure $\mu$, called invariant, is preserved under application of $T$, i.e., $\mu[T^{-1}(A)]=\mu(A)$ for any measurable subset $A$ of the phase space \cite{ER,Hal}. Physical measures are absolutely continuous with respect to the Lebesgue measure, i.e., $d\mu(x)=\rho(x)dx$, being the invariant density $\rho(x)$ an eigenfunction of the Perron-Frobenius (PF) operator $\mathcal{L}$, namely
\begin{equation}
\label{PF}
\mathcal{L}\varrho(x)=\sum_{\xi_{j}=T_{j}^{-1}(x)}\frac{\varrho(\xi_{j})}{|\mathcal{D}T(\xi_{j})|},
\end{equation}
where $\mathcal{D}T$ is the Jacobian determinant of $T$ and the sum is over all its preimages. Among all eigenfunctions $\varrho$ of $\mathcal{L}$, the invariant density is such that $\mathcal{L}\rho=\rho$, whereas the remaining eigenvalues ​​have magnitude less than one.

Characterizing invariant measures, eigenfunctions, and spectra for nonlinear dynamical systems is a fundamental problem which connects ergodic theory with statistical mechanics \cite{ER,Hal}. In the case of Hamiltonian systems, a number of methods have been developed to deal with PF operator, either by resonance spectrum \cite{PRres} or differential operators \cite{RVd}. Despite many advances, it seems still largely impossible to approximate analytically eigenfunctions of the PF operator for a broad class of dynamical systems with dimension more than one. Yet even in the one-dimensional cases, obtaining an invariant measure for a given dynamical system is in general a very difficult task. Methods to solve exactly invariant measures for certain classes or specific interval maps include change of coordinates via topological conjugacy \cite{Ulam}, solvable chaotic maps \cite{Ume}, inverse solutions of the PF operator \cite{EM}, among other possible.

According to Halmos (``unsolved problems'' of Ref. \cite{Hal}), an outstanding problem of ergodic theory is the conjugacy problem: when are two maps  topologically conjugate? Two maps $S$ and $T$ are topologically conjugate if there exists a homeomorphism $\omega$ such that
\begin{equation}
\label{defc}
S\circ\omega=\omega\circ T.
\end{equation}
Of course, the problem refers to the task of finding $\omega$, if it exists, when $T$ and $S$ are given beforehand.

Fundamental aspects of chaotic dynamics are preserved under conjugacy. A map is chaotic if is topological transitive and has sensitive dependence on initial conditions. Both properties are preserved under conjugacy, then $T$ is chaotic if and only if $S$ is also. Moreover, the invariant measure of a map can be obtained by the knowledge of invariant measure of its conjugate. Topological conjugacy gathers different maps into equivalent classes, and therefore helps us understand more complicated dynamical systems in terms of simpler ones. We can mention, for instance, nonlinear flows near a fixed point in terms of the linearised flow: the robustness of such analysis was elucidated by the using of conjugacy in the well-known Hartman-Grobman theorem \cite{Hart}.

The conjugacy problem for interval maps has probably been an even more difficult challenge than solving PF equation exactly. Examples of conjugacy in the literature are remarkably scarce and, with very few exceptions (see for instance Ref. \cite{GT}), almost always boil down to Ulam's example, dating back to 60's \cite{Ulam}. The lack of general methods to identify if a topological conjugacy occurs is therefore a barrier to be overcome.

In this Letter we deal with the conjugacy problem for interval maps taking into consideration the inverse PF problem. The common thread is the knowledge of way whereby monotone partitions of maps are related to each other. The solution of a conjugacy problem between two specific maps does not arise in the form of a specific homeomorphism, but under conditions to be fulfilled by the transformation. Such conditions help either to check that a conjugacy is prohibitive or unveil further maps belonging to the same topological class.

{\it Inverse PF solution} - Let us consider here piecewise monotone maps having finitely many branches $k$. Thus, a given map $T:I\rightarrow I$ is defined on a partition $\{I_{0}, \ldots, I_{k-1}\}$ having full measure in $I$ so that $T|_{I_{j}}:=T_{j}$ are monotone. Let us consider the $k$ branches of such maps in the advantageous form
\begin{equation}
\label{adf}
T_{j}=\Phi_{j}^{-1}\circ\mu,
\end{equation}
being $\Phi_{j}$ monotone functions for all $j$. Let us also introduce the {\it foldings} $h_{j}$ so that $T_{j}=h_{j}\circ T_{0}$, i.e.,
\begin{equation}
\label{fol}
h_{j}=\Phi_{j}^{-1}\circ\Phi,
\end{equation}
where we simply set $\Phi_{0}=\Phi$ and $h_{0}(x)=x$. Equation (\ref{fol}) yields the manner whereby the branches of a map are related. Note that it is necessary to consider here that each branch $T_{j}$ is also a function extending on the whole interval $I$ and with image beyond $I$. Within such framework, after solving Eq. (\ref{PF}) we get the following equation for the invariant measure
\begin{equation}
\label{muvar}
\mu=\Phi+\sum_{j=1}^{k-1}\mbox{sgn}(h'_{j})\Phi\circ h_{j}^{-1},
\end{equation}
where $h'_{j}$ and $h_{j}^{-1}$ denote, respectively, the derivative and the inverse of $h_{j}$, and $\mbox{sgn}$ stands for the sign function.

Thus, by means of the choice of $\Phi$ and $h_{j}$ we are able to construct a map equation with the desired absolutely continuous invariant measure. It is worth to point out that, in the frame of inverse approach, is not trivial that an interval map obtained from an absolutely continuous invariant measure is chaotic, but this is an easily verifiable feature for individual maps.

Although seemingly straightforward, this approach will prove extremely useful for all the results that will follow. We can illustrate its use by means of a well-known unimodal map: the tent map $T(x)=1-|2x-1|$ on $I=[0,1]$ \cite{Ulam}. This model was considered by Lorenz \cite{Lor} as an approximation for the cusp-shaped Poincar\'e first return map in the strange attractor that bears his name. In that case we have $h_{1}(x)=h_{1}^{-1}(x)=2-x$ and $\Phi(x)=x/2$, resulting in the uniform density $\rho(x)=1$. We shall see that the folding approach is quite comprehensive, being even able to describing closed-form (not piecewise-defined) map equations.

{\it Topological conjugacy} - The conjugacy $\omega$ induces a bijection between the space of ergodic invariant measures of $S$ and of $T$: if $\mu$ is an invariant measure for $T$, then the corresponding invariant measure of $S$ is
\begin{equation}
\label{mucj}
\mu_{S}=\mu\circ \omega^{-1}.
\end{equation}
Thus, conjugacy is also very useful because a suitable change of coordinates according to Eq. (\ref{defc}) makes most advantageous determining the invariant measure of a map by means of another. The Ulam-von Neumann logistic map $S(x)=4x(1-x)$ on $J=[0,1]$ and the tent map previously introduced are widely employed as a benchmark of conjugacy, even though its mechanism still remains poorly understood. It is just known that the choice $\omega(x)=\sin^{2}(\pi x/2)$ works, the famous Ulam's example \cite{Ulam}. Ulam knew in advance the invariant densities of both maps, $\rho(x)=1$ and $\rho_{S}(x)=1/[\pi\sqrt{x(1-x)}]$, and may have checked them via equation (\ref{mucj}). However, given the problem of finding explicit solutions of the PF operator for previously established maps, such approach faces clear limitations to be successfully employed as a general method.

We now establish conditions to determine conjugacy of interval maps. Let $T: I\rightarrow I$ and $S:  J\rightarrow J$ be two of such maps preserving measures $\mu(h_{j},\Phi)$ and $\mu_{S}(h_{Sj},\Phi_{S})$, respectively. From now on, without any loss of generality, we will consider $T$ and $S$ rescaled so that $I=J=[0,1]$, unless otherwise noted. Let us also introduce $\epsilon_{j}:=\mbox{sgn}(h'_{j}h'_{Sj})$. If $T$ and $S$ are topologically conjugate, then $\epsilon_{j+1}=\epsilon_{j}:=\epsilon$ for all $1\leq j\leq k-1$ (when $k>2$) and the $\epsilon$-conjugacy of foldings
\begin{equation}
\label{cjmy1}
h_{Sj}\circ\omega=\omega\circ h_{j},\qquad \omega\circ\mathbbm{1}_{\epsilon}=\omega,
\end{equation}
holds for all $j$, being $\mathbbm{1}_{\epsilon}(x):=\epsilon x$. Furthermore, $\Phi$ and $\Phi_{S}$ are such that
\begin{equation}
\label{phini}
\Phi\circ\mathbbm{1}_{\epsilon}=\epsilon\Phi,\qquad\Phi_{S}=\Phi\circ\omega^{-1},
\end{equation}

For the proof first observe that, by comparing each term of $\mu(h_{j},\Phi)$ and $\mu_{S}(h_{Sj},\Phi_{S})$ via Eqs. (\ref{muvar}-\ref{mucj}), Eq. (\ref{phini}) holds together with $\epsilon_{j+1}=\epsilon_{j}$ for all $1\leq j\leq k-1$ (for $k>2$) and $h_{Sj}\circ\omega\circ\mathbbm{1}_{\epsilon}=\omega\circ h_{j}$, from which Eq. (\ref{cjmy1}) is a solution. On the the hand, conjugacy $S_{j}\circ\omega=\omega\circ T_{j}$ leads to $\Phi_{Sj}=\Phi_{j}\circ\omega^{-1}$ via Eq. (\ref{mucj}), and thereby Eq. (\ref{cjmy1}) is mandatory because $\omega\circ h_{j}=\omega\circ\Phi_{j}^{-1}\circ\Phi=\Phi_{Sj}^{-1}\circ\Phi$ and, similarly, $h_{Sj}\circ\omega=\Phi_{Sj}^{-1}\circ\Phi_{S}\circ\omega=\Phi_{Sj}^{-1}\circ\Phi$.

The $\epsilon=-1$ case has some interesting peculiarities. Since $S$ and $T$ are set out on [0,1], the condition $\omega\circ\mathbbm{1}_{\epsilon}=\omega$ can be simply suppressed because any homeomorphism on $[0,1]$ can be trivially rendered as an even function on $[-1,1]$. So what would be the meaning of $\omega\circ\mathbbm{1}_{\epsilon}=\omega$?

Given that $\omega$ is even when $\epsilon=-1$, conjugacy (\ref{defc}) implies that $\omega\circ T$ is also an even function and, therefore, both are invariant under reflection symmetry. This means that $\omega\circ T(x+1)=\omega\circ T(x)$, and thus $S\circ\omega(x)$ on $[0,1]$ is identical to $\omega\circ T(x+1)$ on $[-1,0]$. Assuming $\omega$ as a periodic function throughout the real line starting from the cell $[-1,1]$ it is easy to see that, out of $[0,1]$, the same applies to $T(x+N)$ for every nonzero integer $N$. It has also not escaped our notice that the $\epsilon-$conjugacy is not invariant under exchanging $T\leftrightarrows S$, though the original conjugacy (\ref{defc}) itself is via $\omega\leftrightarrows\omega^{-1}$. This can be checked by noting that $\omega=\Phi_{S}^{-1}\circ\Phi$. Since $\epsilon=-1$ implies that $\Phi$ must be an odd function, $\omega\circ\mathbbm{1}_{\epsilon}=\omega$ leads to $\Phi_{S}^{-1}\circ\Phi=\Phi_{S}^{-1}\circ\Phi\circ\mathbbm{1}_{\epsilon}=\Phi_{S}^{-1}\circ(-\Phi)$, therefore prohibiting $\Phi_{S}$ of being also an odd function. What we have got here is a case of spontaneous symmetry breaking, where solutions of an equation have a lower symmetry than the equation itself \cite{PGs}. This is explained by the existence of further solutions $T (x + N)$, among other possible, that also satisfy $S\circ\omega=\omega\circ T(x+N)$, but with noninvertible $\omega$ beyond $[0,1]$.

If $\Phi$ is an odd function (regardless of $\epsilon=\pm1$), then conjugacy leads to some specificities for $S$ and $T$ due to the form of Eq. (\ref{adf}). By Eq. (\ref{phini}), $\Phi$ monotone is such that $\Phi(0)=0$, and thus $T(0)=\Phi^{-1}\circ\mu(0)=0$. Moreover, $S_{0}\circ\omega=\Phi_{S}^{-1}\circ\mu$, hence $S_{0}(0)=\Phi_{S}^{-1}(0)$. From Eq. (\ref{phini}), $\Phi_{S}$ monotone is such that $\Phi_{S}\circ\omega(0)=\Phi(0)=0$, resulting $S(0)=T(0)=0$ for conjugate maps.

The framework (\ref{cjmy1}-\ref{phini}) and details thereof are well ilustrated by cosidering again the tent and Ulam-von Neumann maps, for which $h_{1}(x)=2-x$ and $h_{S1}(x)=x$, respectively. Thus we have $\epsilon=-1$, $\Phi(x)=x/2$, and $\omega$ satisfying the periodicity condition $\omega(x)=\omega(2-x)$, from which $\omega(x)=\sin^{q}(\pi x/2)$ on $[0,1]$ is a family of solutions for $q>0$, yielding $S(x)=2^{q}x(1-x^{2/q})^{q/2}$. Thus, the Ulam-von Neumann map comes from $q=2$, but others solutions are also possible. For example, Jacobi's sn elliptic function is a doubly periodic generalization of sine function, also fulfilling the same periodic criteria for $\omega(x)=\mbox{sn}^{2}(Kx,\kappa)$, where $K(\kappa)$ is the complete elliptic integral of the first kind and $\kappa$ its elliptic modulus \cite{AS}. By considering the addition identity of sn function, this choice gives us Schr\"oder's map $S(x)=4x(1-x)(1-\kappa^{2}x)/(1-\kappa^{2}x^{2})^{2}$ \cite{Scp}. After Schr\"oder, this map was also considered by Latt\`es and, recently, by Milnor \cite{Milnor}, among other authors \cite{Ume}.

We can also check examples in the literature: the map $S(x)=\sqrt[3]{1/8-2|x-1/2|^{3}}+1/2$ proposed in Ref. \cite{GT} gives $h_{S1}(x)=1/2+\sqrt[3]{1/4-(x-1/2)^{3}}$ and $\epsilon=1$ when compared with the tent map. Conjugacy of foldings yields $\sqrt[3]{1/4-(\omega-1/2)^{3}}=\omega(2-x)-1/2$, and therefore $f(2-x)=1/4-f(x)$ by taking advantageously $\omega(x)=1/2+\sqrt[3]{f(x)}$. Thus, the linear solution $f(x)=(1/4)(x-1/2)$ fulfils conjugacy between both maps. Of course, further solutions for $f$ that keep $\omega$ as a homeomorphism generate different conjugate maps for $S(x)$.

A simple example, though very illustrative, is the case of Bernoulli shift map (also called dyadic transformation) $T(x)=2x$ mod $1$. Would this map topologically conjugate to the Ulam-von Neumann map $S(x)$? For the Bernoulli map we have $h_{1}(x)=x-1$ and $\Phi(x)=x$, resulting $\epsilon=1$ and $\omega(x)=\omega(x+1)$. Evidently, there is no homeomorphism $\omega$ on $[0,1]$ such that $\omega(0)=\omega(1)$. However, $\omega(x)=\sin^{2l}(\pi x)$ fulfills conditions above for $l\geq1$ integer, and it is not difficult to see that $S\circ\omega=\omega\circ T$ for $S(x)=2^{2l}x(1-x^{1/l})^{l}$. The $l=1$ Ulam-von Neumann map is a well-known case of so-called semi-conjugacy, when $\omega$ is noninvertible. This example behaves like a $\epsilon=-1$ case, where $\Phi$ is odd and $\omega$ is even, having thus symmetry breaking.

Notwithstanding the example above, we can notice that $\omega$ does not need to be restricted to a homeomorphism. In fact, since we want obtain and relate absolutely continuous invariant measures, we expect that $\omega$ is a piecewise diffeomorphism. Then we can ask whether there is any map $S(x)$ topologically conjugate to the Bernoulli shift map, and also having the same neutral folding of the Ulam-von Neumann map. There is an invertible transformation $\omega$ satisfying $\omega(x)=\omega(x+1)$, namely $\omega(x)=-\cot(\pi x)$, yielding the rational map  $S(x)=(1/2)(x-1/x)$ on the real line. Interestingly, this equation is the Newton-Raphson map for the roots of equation $x^{2}+1=0$. Since there is no real root, the iterations of the map do not converge to a limit and its behavior is therefore chaotic. But more importantly, note that we have here $S\circ\omega_{N}=\omega_{N}\circ T(x+N)$ where, for each integer $N$, $\omega_{N}: [N, N+1]\rightarrow(-\infty,\infty)$. It is a single conjugacy problem $\omega_{N}$ for each pair $S(x)$ and $T(x+N)$, and the symmetry breaking therefore does not occur. This case, together with the previous example, illustrates why $\epsilon=1$ is not a decisive symmetry breaking parameter.

\begin{figure}[h]
\centering
\includegraphics[scale=0.20]{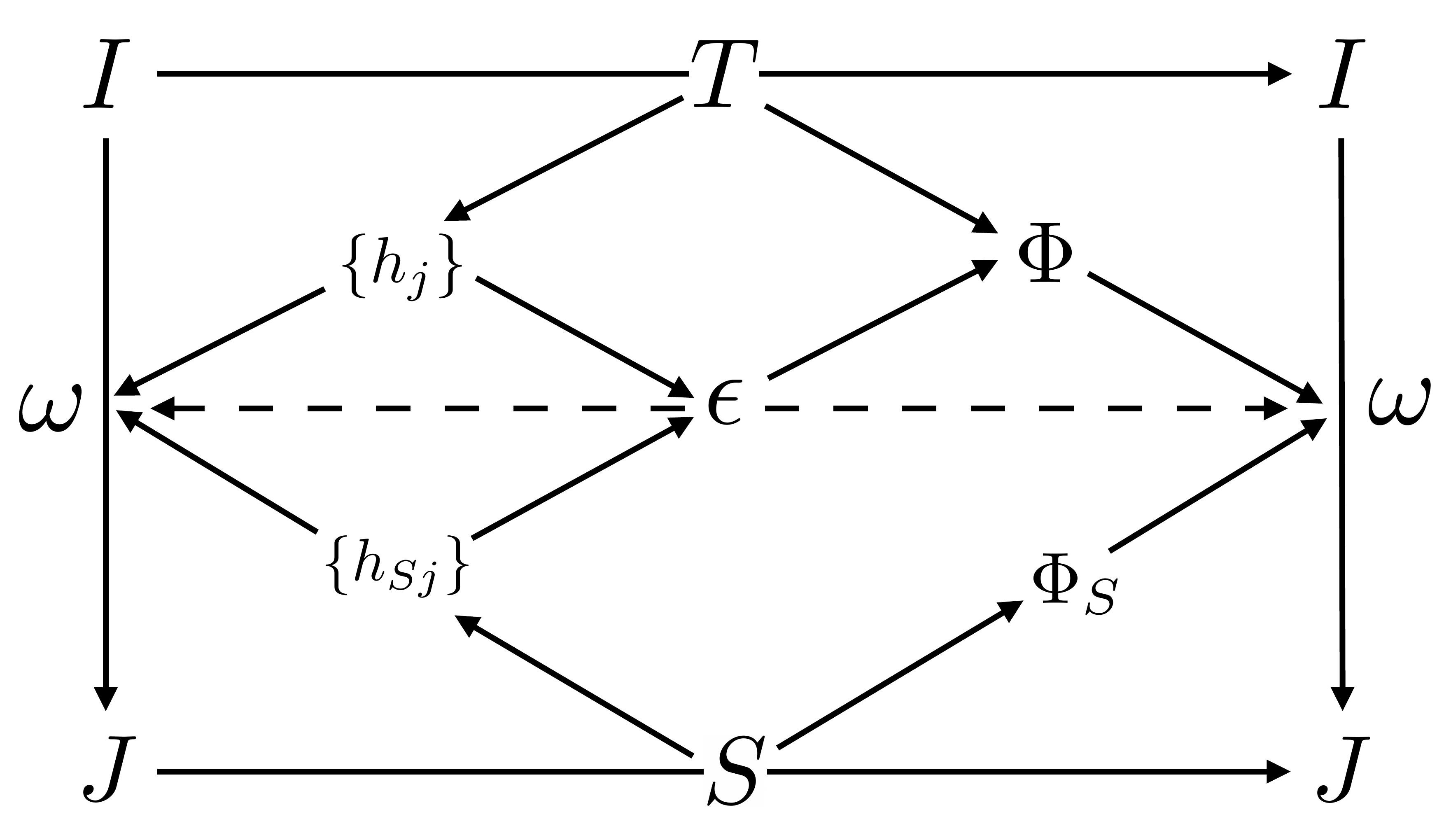}
\caption{Topological conjugacy is usually represented by a commutative diagram, like the rectangle with vertices $I$ and $J$ depicted above. Here, such diagram is supported by the symmetry-breaking framework (\ref{cjmy1}-\ref{phini}), highlighting the feasible paths for the conjugacy problem of interval maps.}
\label{fig1}
\end{figure}

A commutative diagram depicting all steps of the conjugacy problem is showed in Fig. \ref{fig1}.

{\it Exploring eigenfunctions of the PF operator} - We can also investigate the relationships between eigenfunctions and spectra of conjugate maps. Let $u_{\sigma}$ be the antiderivative of an eigenfunction with corresponding eigenvalue $\sigma$, i.e., $u'_{\sigma}=\varrho_{\sigma}$. Let us introduce $\tilde{h}_{j}:=T_{0}^{-1}\circ h_{j}\circ T_{0}$, where $T_{0}=\Phi^{-1}\circ\mu$ extends throughout $I$. By means of the PF operator we get the following relation
\begin{equation}
\label{bern}
\sigma u_{\sigma}\circ T_{l}=u_{\sigma}\circ\tilde{h}_{l}+\sum_{j=1}^{k-1}\mbox{sgn}(\tilde{h}'_{j})u_{\sigma}\circ\tilde{h}^{-1}_{j}\circ\tilde{h}_{l},
\end{equation}
for $0\leq l\leq k$. We also observe that, it $T$ and $S$ are conjugate, then the pair $(\tilde{h}_{j},\tilde{h}_{Sj})$ is also conjugate. By considering this property and applying $\omega^{-1}$ on the left side of Eq. (\ref{bern}), the antiderivative of eigenfunctions  and corresponding spectra of the two conjugate maps are such that
\begin{equation}
\label{spec}
u_{\sigma}\circ\mathbbm{1}_{\epsilon}=\epsilon u_{\sigma}
,\qquad u_{S\gamma}=u_{\sigma}\circ\omega^{-1},\qquad \gamma=\sigma.
\end{equation}

Owing to the nature of topological conjugacy, the result $\gamma=\sigma$ is not unexpected, but the parity property and the way in which $u_{S\gamma}$ and $u_{\sigma}$ are transformed, just as in Eq. (\ref{phini}), do not seem to be available or hinted in the literature as far as I know. Of course, the knowledge of a certain map spectrum lets us access to the spectra of all corresponding conjugate maps. Otherwise, obtaining a map spectrum is tipically a difficult task, being very rare analytical solutions, see for instance Ref. \cite{PRres} and references therein. Notwithstanding, Eq. (\ref{bern}) becomes more handleable for neutral foldings, i.e., $h_{j}(x)=x$ for all $j$, where one has the solvable form
\begin{equation}
\label{solv}
\Phi\circ T=k\Phi
\end{equation}
via Eqs. (\ref{adf}) and (\ref{muvar}). The semigroup property of solvable maps (\ref{solv}) enable us the acquaintance of any iterations in just one computable step despite the map being chaotic, manny examples are discussed in \cite{Ume}. The Ulam-von Neumann map is also the well-known example of exactly solvable chaotic map, whose iterations follow $x_{n}=\sin^{2}(2^{n}\arcsin\sqrt{x_{0}})$, recalling that Eq. (\ref{phini}) yields $\Phi_{S}(x)=(1/2)\omega^{-1}(x)=(1/\pi)\mbox{arcsin}\sqrt{x}$. Similar relations also holds for Schr\"oder's map, in this case $\omega(x)=\mbox{sn}^{2}(Kx,\kappa)$, among others maps. Such kind of solvable chaotic dynamics has been generated by means of optical devices composed of multiple Mach-Zehnder interferometers \cite{UAK}. Moreover, their potential as platforms for chaos-based public-key cryptography have also been considered, see Ref. \cite{MK} and references therein.

For exactly solvable chaotic maps (\ref{solv}), Eq. (\ref{bern}) becomes $u_{\sigma}\circ T=(k/\sigma)\,u_{\sigma}$, i.e., $u_{\sigma}$ and $\Phi$ are solutions of nonidentical Schr\"oder's functional equations \cite{Scp}. Thus, the solution for $u_{\sigma}$ is such that $u_{\sigma}=\psi_{\sigma}\circ\Phi$ and $\psi_{\sigma}(kx)=(k/\sigma)\psi_{\sigma}(x)$. By setting $\sigma=k^{-\gamma}$, we finally get
\begin{equation}
\label{usig}
u_{\gamma}(x)\propto\Phi^{1+\gamma}(x),\qquad \gamma\geq0,
\end{equation}
which corresponds to $\varrho_{\gamma}\propto\Phi^{\gamma}\Phi'$. In order to compare our results with those in the literature, let us consider $u_{S\gamma}\propto\Phi_{S}^{1+\gamma}$ for the Ulam-von Neumann map and its conjugacy $\omega$ with the tent map. Our Eq. (\ref{spec}) yields $\varrho_{\gamma}(x)\propto(x/2)^{\gamma}$ with $\gamma=2l$ for the tent map, being $l$ a nonnegative integer. Since $\mathcal{L}$ is a linear operator, then we can use Bernoulli polynomials as eigenfuntions, i.e., $\varrho_{l}(x)=B_{2l}(x/2)$. Of course, $B_{2l+1}(x/2)$ is the null space (with $\sigma=0$). This result is in full agreement with those in Ref. \cite{Dorf}.

{\it Concluding remarks} - We proposed a decomposition of maps based on how their monotone branches fold on each section of phase space. Such characterization enables us to shift the complexity of solving PF operator to the form of function $\Phi$, which is very advantageous when employing an inverse approach. Thus, our results also helps to understand the rareness of obtaining the invariant measure expression explicitly from a given map, which would be naively natural way to get it.

Within such approach, we develop a method of determining topological conjugacy of interval maps, the main result of this manuscript. Identify such property has been mostly a kind of art, where the reasons behind a conjugacy show up invariably hazy in the literature. Besides highlighting something deeper on the relationship between branches of maps, the connection between the conjugacy and PF problems proved to be very fruitful. The method was successfully checked beyond the maps that are commonly used as benchmark studies of conjugacy. The eigenfunctions of the PF operator for maps under conjugacy are such that $u_{\sigma}$ and $\Phi$ follow exactly the same transformation laws, enabling to relate the spectra and eigenfunctions in a quite straightforward way. In the case of solvable chaotic maps we provide a complete solution of the PF problem, covering the invariant measure, eigenfunctions and spectrum.

We have seen here that the conjugacy problem reveals itself an intrinsically non-local problem, with existing maps also outside of intervals previously conceived. The original relationship (\ref{defc}) does not distinguish {\it a priori} a pair of genuinely conjugate maps from further (external) maps conjugate by noninvertible transformations. Our approach signals the existence of such multiplicities by means of spontaneous symmetry breaking, because the non-commutability of $S$ and $T$ via conjugacy just takes place when $\omega$ is noninvertible. Off course, semi-conjugate solutions fall in this same scenario. Conjugacy also has the transitive property, where a closed chain of conjugate maps are relate to each other through different transformations. Thus, it seems reasonable that conjugacy criteria should be established by means of general properties of maps, enabling to gather the largest possible number of elements in such a chain, rather than get a only pair of elements. Indeed, this is perfectly consistent with the fact that conjugacy means topological equivalence of all maps belonging to a same chain. We believe that the method presented here fits in such a frame.

{\it Acknowledgements} - The author thanks Alberto Saa for valuable discussions. This work was supported by Conselho Nacional de Desenvolvimento Cient\'ifico e Tecnol\'ogico (CNPq), Brazil (Grant No 307618/2012-9) and special program PROPES-Multicentro
(UFABC), Brazil.

\end{document}